\documentclass[12pt]{iopart}

\usepackage{graphicx}
\usepackage{amssymb}

\begin{document}

\newcommand{\figuresize}{0.90\columnwidth}
\newcommand{\doublefiguresize}{1.95\columnwidth}

\title{Exploration of Order in  Chaos with Replica Exchange Monte Carlo}

\author{Tatsuo Yanagita}
\address{Research Institute for Electronic Science, Hokkaido University, Sapporo 001--0020, JAPAN}
\ead{yanagita@nsc.es.hokudai.ac.jp}

\author{Yukito Iba}
\address{The Institute of Statistical Mathematics, Tokyo 106--8569, JAPAN}
\ead{iba@ism.ac.jp}

\begin{abstract}
A method for exploring unstable
structures generated by nonlinear dynamical systems
is introduced. It is based on the sampling of initial
conditions and parameters by 
Replica Exchange Monte Carlo (REM),
and efficient both for the search of rare initial 
conditions and for the combined search of 
rare initial conditions and parameters.
Examples discussed here include the sampling
of unstable periodic orbits in chaos and search for
the stable manifold of unstable fixed points,
as well as construction of the global bifurcation 
diagram of a map.
\end{abstract}

%%%%%%%%%%%%%%%%%%%%%%%%%%%%%%%%%%%%%%%%%
\section{Introduction}

Nonlinear dynamical systems exhibit very complex structures 
in the phase space~\cite{Ott02,CvitanovicWebBook}. 
There can be a number of stable or unstable 
fixed points, periodic orbits, chaotic saddles, and more
intricate invariant manifolds.
Basin structures corresponding to these invariant
manifolds can also be very complicated. Quest for 
these structures is an important subject in the study
of dynamical systems.

It is, however, not easy to capture these global structures 
by time evolution from randomly chosen initial conditions, 
because they are often unstable and correspond to
very rare initial conditions. Analysis of these structures
requires an efficient algorithm for sampling rare trajectories
from rare initial conditions, which acts as an anatomical tool 
of nonlinear dynamical systems.
Moreover, properties of chaotic or 
multi-basin dynamical 
systems are often sensitive to
the choice of values of the parameters. 
Then, additional complexity arises when we do not 
have precise knowledge on the values of parameters 
with which interesting structures appear.
Algorithm with which we can perform
combined search of the space of initial conditions
and of the parameter space is required for 
analyzing dynamical systems.

In this paper, we introduce a method based on 
Markov Chain Monte Carlo (MCMC)~\cite{Landau05,Newman99,Liu01}, which 
is efficient both for the search of rare initial conditions and for the
combined search of rare initial conditions and parameters.
In the proposed algorithm, we identify 
initial conditions of trajectories as microscopic states
to be sampled,  and assign each of them an artificial
``energy'' that represents
a measure of  ``atypicality'' of the trajectory 
starting from the initial condition.
These initial conditions are sampled by MCMC and
a set of atypical trajectories is obtained. 
This approach is easily extended to the
combined search of the space of initial conditions
and of the parameter space. An essential idea is
the use of direct product of initial 
conditions and parameter as a state space explored
by MCMC. 
 
In both cases, however, naive applications
of MCMC will be impaired by sensitive dependence on
initial conditions and
parameters in nonlinear dynamical systems, which 
lead to a highly multimodal energy function. 
Here we use {\it Replica Exchange Monte Carlo} (REM), which is
also known as {\it Parallel Tempering} (PT), to circumvent
the difficulty.
REM is useful for the sampling
on a rugged energy surface and intensively
used for the finite-temperature simulation of spin glass~\cite{Young97,Janke08}
and biomolecules~\cite{Mitsutake01,Janke08}. 
When we generate multiple samples with 
conventional optimization methods, such as simulated
annealing, it is difficult to control the probability of
repeated appearance of the same (or similar) items
in an obtained set of samples. 
The advantage of REM is
that it realizes unbiased sampling from the given ensemble
even in the region corresponding to low ``temperatures''.
%This is the reason that REM is often preferred
%for the finite-temperature simulation of spin glass (refs)
%and biomolecules (refs.). 
%In the present context,
%it is useful to study diversity and probability of 
%the trajectories with the desired property. 

The idea of using MCMC or related algorithm 
for sampling  rare trajectories in nonlinear dynamical
systems  is already seen in 
literature~\cite{Cho94,Bolhuis98,Vlugt00,
Kawasaki05,Sasa06,Giardin06,Tailleur07}, while
combined search of initial conditions and parameters
is rarely studied. 

These studies are classified into two categories. 
Some of them~\cite{Cho94,Bolhuis98,Vlugt00,
Kawasaki05,Sasa06}, including an earlier 
work~\cite{Cho94} and recent studies~\cite{Sasa06,Kawasaki05}, 
developed frameworks where the states to be sampled by
the algorithm are {\it entire trajectories} or {\it
orbits}, instead of {\it initial conditions} of the trajectories.
The ``transition path 
sampling''~\cite{Bolhuis02,Bolhuis98,Vlugt00,vanErp07}, which 
mostly used for sampling trajectories between two 
metastable states, is
also based on a similar choice of state space, that
is, a state consists of the array of positions 
(and momentums) of particles at all time-steps in
a trajectory. 
The algorithm proposed 
in this paper is much 
simplified with the choice of initial condition 
as state variable. It is also a natural choice for 
a deterministic but not necessary reversible dynamical system. 
In the following sections, we will show that 
the proposed algorithm can deal with impressively
various kinds of problems.  

Another type of algorithm proposed 
in~\cite{Giardin06,Tailleur07}
approximates ``pseudo trajectories'' with desired
property by a set of particles each of which 
obeys original equation of motion. 
Using a genetic algorithm like
split/delete procedure, efficient sampling of
rare trajectories is realized. It is a kind of sequential Monte Carlo~\cite{Doucet01,Iba01p}
or diffusion Monte Carlo~\cite{Kalos62,Iba01p}
and not genuine MCMC.
It can also be interpreted as  a multi-particle 
version of ingenious ``staggered-step'' 
algorithm~\cite{Sweet01} 
developed earlier. The idea seems
effective for trajectories with positive Lyapunov numbers, but
its advantage might be reduced in a search for 
trajectories with negative Lyapunov numbers where split
trajectories does not diverge fast without strong 
external noise.
Anyway, it seems to capture different aspects of chaotic systems
and will be complementary to the method proposed 
in this paper.

The organization of the paper is as follows:
In Section.2, we explain the basics of the proposed
method. In Section.3, two examples of initial
condition sampling are discussed. The first one
is a toy example of the sampling of the 
unstable periodic orbits
of the Lorenz model. The second one is
sampling of stable manifold of the unstable 
fixed points of a double pendulum with dissipation, which is a much
difficult and interesting example. In Section.4,
search of parameter space and 
combined search of parameter space and
initial condition space with a modified algorithm are studied.
Section.5 devoted to the discussion on 
the results and possibility of
further extensions.

%%%%%%%%%%%%%%%%%%%%%%%%%%%%%%%%%%%%%%%%%
\section{Method}
\label{sec:method} 

\subsection{State Vector}

In the basic algorithm for the discrete
time dynamics with a given set of parameters,
the initial condition $X=(x_1,...,x_n) \in \mathbb{R}^n$
of a trajectory is used as a state variable 
that characterizes the trajectory evolving
from the initial condition. 
In a continuous time case, it is often better to include
the time $T$ when desired phenomena occur. Then a
state will be $X=(x_1,\ldots,x_n,T) \in \mathbb{R}^{n+1}$.
Finally, when the parameters are also unknown, 
the state vector will be 
$X=(x_1,\ldots,x_n,T,\alpha_1,\ldots,\alpha_m) \in \mathbb{R}^{n+1+m}$,
where $\alpha=(\alpha_1,\ldots,\alpha_m)$ is a vector of unknown
parameters.

%%%%%%%%%%%%%%%%%%%%%%%%%%%%%%%%%%%%%%%%%
\subsection{Energy Function and Gibbs Distribution}

In order to explore rare structures in phase space, we 
should construct a fictitious  ``energy'' function 
$E(X)$ defined on the state space.
The function depends on the kind of rare orbit we want to explore.
For the detection 
of periodic orbits of a map $x_{n+1}=f(x_n)$,
a simply possibility of the 
energy function is $E(x_0)=\log(|f^n(x_0)-f(x_0)|)$, where the state
$X$ coincides with  
the initial condition $x_0$ of the iteration. 
The other choices of the energy function to explore atypical structure
in phase space are shown in Sec.~\ref{sec:extension}, where
$E$ may depend also on $T$ and $\alpha$. 

Once we define the energy function, it is straightforward to 
define the Gibbs distribution with the energy
\begin{equation}
p(X)=\frac{\exp(-\beta E(X))}{Z(\beta)}{}, 
\qquad Z=\int \exp(-\beta E(X)) dX {}_. \label{Gibbs}
\end{equation}
While $p(X)$ coincides with the uniform density in a prescribed
region when $\beta$ takes a small value, it concentrate to 
regions with small values of $E$ when $\beta$ becomes large,
where we can collect samples of
trajectories (and parameters) of desired properties.  

%%%%%%%%%%%%%%%%%%%%%%%%%%%%%%%%%%%%%%%%%
\subsection{Metropolis Algorithm}

Now, the problem of sampling atypical trajectories 
is reduced to the sampling from the density $p(X)$. 
Here we use the Metropolis 
algorithm~\cite{Metropolis53}, a simplest implementation
of the idea of MCMC.
The Metropolis algorithm used here is
essentially the same as the standard one commonly used in
statistical physics. There is, however, an 
important difference in actual implementation, that is,
we should simulate the trajectories of the length $T$ 
from the initial
condition $x$ with the parameter $\alpha$ at each trial
of changing $x$, $T$, and $\alpha$. 
If we explicitly represent this difference, the iteration of 
algorithm is described as follows.

\begin{enumerate}
\item 
Draw a perturbation to current states 
$(\triangle x, \triangle T, \triangle \alpha)$ 
from a prescribed  ``trial'' density $q$, which
 defines a move set. 
Hereafter, the mirror symmetry 
\[
q(-\triangle x, -\triangle T, -\triangle \alpha)
=q(\triangle x, \triangle T, \triangle \alpha)
\]
of the density $q$ is assumed.
Then, when the current values of 
$(x, T, \alpha)$ is 
$(x^{(n)},T^{(n)},\alpha^{(n)})$, 
the candidate of 
the next states is defined as 
$(x',T',\alpha')=(x^{(n)}+\triangle x,T^{(n)}+\triangle T,
\alpha^{(n)}+\triangle \alpha)$.
%In the following examples, we choose $q$ that tries to change
%one and only one component of $(x,T,\alpha)$
%at a time. 
\label{move}

\item
Run the simulation of length $T'$ 
of the dynamical system with parameter $\alpha'$ 
from the initial condition $x'$. From the 
result, $E(\alpha',T',x')$ is computed.  

\item 
Draw a uniform 
random number $R \in [0,1]$. If and only if
$$
R<\frac{\exp(-\beta E(x',T',\alpha'))}
{\exp(-\beta E(x^{(n)},T^{(n)},\alpha^{(n)}))} {} \ ,
$$
the new proposal is accepted: 
$
x^{(n+1)}=x', \, T^{(n+1)}=T', \, \alpha^{(n+1)}=\alpha' {}_.
$
Else nothing is changed:
$
x^{(n+1)}=x^{(n)}, \, T^{(n+1)}=T^{(n)}, \, \alpha^{(n+1)}=\alpha^{(n)}
$

\item Return to Step~(i).
\end{enumerate}

An important point for treating unstable structures in
potentially chaotic systems is the choice of the density 
$q$, or, equivalently,
the set of moves. The point is that 
the moves should be hierarchical for
the efficient sampling, that is, the coexistence of 
tiny and large changes in phase space is essential~\cite{Sweet01}.
%The large moves are
%useful for global search and the tiny moves are
%necessary to adjust special initial condition with high precision.
Here we adopt the method introduced in~\cite{Sweet01}, that is, 
the elements $\triangle x_i$ of the perturbation
$\triangle x$ is given by $\delta x_i= s \times d \times 10^{-e}$
where $d$ and $e$
are random integers uniformly distributed
in $d \in [1,9]$ and $e \in [N_e^{\min},N_e^{\max}]$,
and $s$ is a binary random number that takes the value
$\pm 1$ with probability 0.5, respectively.
The parameters $N_e^{\min}$ and $N_e^{\max}$ 
of the algorithm define
the logarithmic scales of the largest and smallest perturbations.
The corresponding trial density $q(x_i)$ becomes
a mixture of uniform distributions with the different scales. 
It has a sharp peak near zero as well as a very long tail.
The components of $\triangle\alpha$ and $\triangle T$
are also generated by a similar way. 
%with possibly 
%different values of $N_e^{\min}$ and $N_e^{\max}$.

%For the definition of the moves,
%it is useful to introduce a M-ary representation
%as follows. Each element $x_i$ (or $T$, $\alpha_j$) of the state 
%vector is a finite precision number in a computer
%and can be represented as
%%
%In the context of 
%statistical physics, $\{x_{ik}\}$ can be regarded as
%Ising model like representation of initial condition.
%\begin{equation}
%x_i= \sum_{k=-K_{\min}}^{K_{\max}} x_{ik} \, M^{k} {}_, \qquad
%T= \sum_{k=-K'_{\min}}^{K'_{\max}} T_{k} \, M^{k} {}_, \qquad
%\alpha_j= \sum_{k=-K''_{\min}}^{K''_{\max}} \alpha_{jk} \, M^{k} {}_,
%\label{bit}
%\end{equation}
%where the variables $x_{ik}$ and $T_k$, $\alpha_{jk}$ are assumed 
%to take the values of $0$ or $1$.

%The constants such as $K_{\min}$ and $K_{\max}$ define the precision and 
%the largest number in the present representation of real number. 
%We will see that this representation provides
%a convenient way to describe the set of Metropolis moves
%used in the algorithm.
%
%Here we use a method introduced in the ref, which can be 
%well described with a binary representation $(x_{ik},T_k,\alpha_{jk})$ 
%of $X=(x,T,\alpha)$ defined in (\ref{bit}).
%For simplicity, we explain the case where the state variable does not
%contains $T$ and $\alpha$, the step~\ref{move} of the algorithm
%is represented as
%\begin{enumerate}
%\item Choose $k$ uniformly in $\{K_{min} \dots K_{max}\}$
%\item Define $x'_i$ by flipping the $k_{th}$ bit  of $x$: 
%$x'_{ik}=1-x_{ik}$ and $x'_{ik}=1-x_{jk}$ ($j \neq j$).
%\end{enumerate}

This version of Metropolis
algorithm appears to provide a simple and universal way of 
treating the Gibbs distribution~(\ref{Gibbs}). 
The efficiency of 
the algorithm, however, can be reduced when $\beta$ become
large in the case of a highly multimodal energy function. 
This difficulty will be
treated by replica exchange Monte Carlo algorithm 
described in the next subsection.

%%%%%%%%%%%%%%%%%%%%%%%%%%%%%%%%%%%%%%%%%
\subsection{Replica Exchange Monte Carlo}

Replica exchange Monte Carlo (REM) provides 
an efficient way to investigate systems with rugged free-energy 
landscapes \cite{Hukushima96,Iba01,Janke08}, particularly at low temperatures. 
In the present context, it is used in 
references~\cite{Kawasaki05,Sasa06}
for the sampling of unstable orbits.
It is also used in the framework of ``transition path
sampling'' in references~\cite{Vlugt00,vanErp07}.

In a replica exchange Monte Carlo simulation, 
a number of systems $\{X_{(m)} \}$ with different inverse 
temperatures $\beta_m$ ({\it replicas}) are simulated in parallel. 
At regular intervals, an attempt is made to exchange the
configurations of selected, usually adjacent, pair of 
replicas. It is accepted with the probability 
$$
P(X_{(m+1)} \leftrightarrow X_{(m)})=\min[1,\exp(\Delta \beta \Delta E)],
$$
where $\Delta \beta=\beta_{m+1}-\beta_m$ is the difference between the inverse 
temperature of the two replicas and $\Delta E=E(X_{(m+1)})-E(X_{(m)})$ is
the energy difference of them. 

The exchange of replicas with different temperatures
effectively reproduces repeated heating 
and annealing, which avoids trapping 
in local minima of the energy. 
Note that it is especially useful with hierarchical
moves defined in the previous 
subsection, because large moves are accepted at high temperatures
and tiny moves are dominated at low temperatures.

On the other hand, the above rule of stochastic exchange preserves
the joint probability distribution
$$
\prod_m P_{\beta_m}(X_{(m)}) 
$$ 
as shown in the literature, {\it e.g.,}~\cite{Iba01}.
Thus, even with the annealing effect, the probability distribution
of each replica $X_{(m)}$ coincides with $P_{\beta_m}
(X_{(m)})$, which means that an unbiased set of samples is obtained at all 
inverse temperatures $\{\beta_m\}$.

%%%%%%%%%%%%%%%%%%%%%%%%%%%%%%%%%%%%%%%%%
\section{Initial Condition Sampling}

In this section, we give examples of 
the initial state sampling by the proposed 
method.
An example is the search for unstable periodic orbits (UPOs)
of the Lorenz model.
Another interesting example is sampling of 
rare orbits in a double pendulum 
system, i.e., initial conditions 
that locate on the stable manifold of unstable fixed points 
are sampled by the proposed method.

%%%%%%%%%%%%%%%%% Lorenz %%%%%%%%%%%%%%%%%%%
\subsection{Unstable Periodic Orbits of the Lorenz model}

In this subsection, we show that unstable periodic 
orbits (UPOs) in a continuous-time dynamical 
system can be detected by the proposed method. 
UPOs are considered as
important objects that govern the properties
of chaos~\cite{CvitanovicWebBook} and there are considerably many works
that deal with the computation of UPO~\cite{Biham89, Diakonos98,
Davidchack99, Lan04, Kawasaki05, Sasa06}. 
Our purpose here is to show that how the proposed 
method works with a familiar example, but not to prove that the proposed
method is superior to all of these existing methods. 
It will be, however, useful to note that the proposed method
with REM can generate UPOs uniformly
under the assumption of uniform measure in the space of initial
conditions. This suggests that it can be useful for the global
search for UPOs in combination with other methods.

Here, we consider the Lorenz equations \cite{Lorenz63,Sparrow82}
\begin{equation} \label{eq:Lorenz}
\left\{
\begin{array}{ll}
\dot{x}=\sigma(y-z)\\
\dot{y}=rx-y-xz\\
\dot{z}=xy-bz,
\end{array}
\right.
\end{equation}
as a typical continuous-time chaotic system,
where $\sigma, r$ and $b$ are parameters of the system.
The state of the system is denoted in a vector format as
$\vec{x}=(x,y,z)$ and the initial condition is written
as $\vec{x}_0=(x_0,y_0,z_0)$. The flow generated by the
dynamical system~(\ref{eq:Lorenz}) is expressed as $\varphi^t$, and
the orbit determined by the initial condition $\vec{x}_0$ is written
as $\vec{x}_t=\varphi^t(\vec{x}_0)$.

The state sampled by REM is defined 
by $X \equiv (\vec{x}_0,T)=(x_0,y_0,z_0,T)$. Then, 
a candidate of the energy function is given as
\begin{equation} \label{eq:lorenz_energy1}
E(\vec{x}_0,T) 
= \log(|\varphi^T(\vec{x}_0)-\vec{x}_0|+\epsilon)  .
\end{equation}
The parameter $\epsilon \ll 1$ is a constant
for avoiding the divergence of the energy. 
When the orbit $\{\varphi^t(\vec{x}_0): t \in \mathbb{R}\}$ is periodic,
there exists a time $T$ such that $\varphi^T(\vec{x}_0)=\vec{x}_0$, and
the energy of the initial condition $\vec{x}_0$ takes 
the minimum value $\log(\epsilon)$.

There are, however, problems 
with the naive choice (\ref{eq:lorenz_energy1}) of the energy function.
First, the energy $E(\vec{x}_0,0)$ always 
takes the minimum value, because $\varphi^0(\vec{x})=\vec{x}$.
Also, when an initial state $\vec{x}_0$ locates on a fixed point, 
the energy of the initial state takes the minimum value for all $T$.
This implies that the almost all initial conditions
sampled by the above energy function will be 
in the vicinity of fixed points. 
To avoid these unfavorable situations, we will add a 
penalty term $P(\vec{x}_0,T)$ 
to the naive energy function~(\ref{eq:lorenz_energy1}).
An improved energy function is
\begin{eqnarray}
E(\vec{x}_0,T) 
&=& \log(|\varphi^T(\vec{x}_0)-\vec{x}_0|+P(\vec{x}_0,T)+\epsilon)  \\
\label{eq:lorenz_energy2}
P(\vec{x}_0,T) &=& g \left( \frac{1}{T}\int_0^T 
{\varphi^\tau(\vec{x}_0)}^2 d\tau,v_c \right )+g(T,T_c),
\label{eq:lorenz_penalty}
\end{eqnarray}
where we use an auxiliary function
$g(s,s_c)  =  \Theta(s_c-s)(\frac{1}{s}-\frac{1}{s_c})$.
$\Theta(s)$ is the Heviside function defined by
$\Theta(s)=1$ if $s \geq 0$ else $\Theta(s)=0$. 
The first term in the equation (\ref{eq:lorenz_penalty}) 
represents a penalty for slower ``average speed'' of the trajectory,
where $v_c$ is a threshold parameter. 
When an initial condition is in the vicinity of 
a stable or unstable fixed point, the orbit stays 
near the point within a certain time. Then, the averaged
speed becomes slower and the value of integral
in the penalty term becomes smaller, which causes 
a large value of the energy.
The second term is a penalty to the 
states that have small $T$, where
$T_c$ is a threshold parameter.

\begin{figure*}[ht]
\begin{center}
\resizebox{\figuresize}{!}{\includegraphics{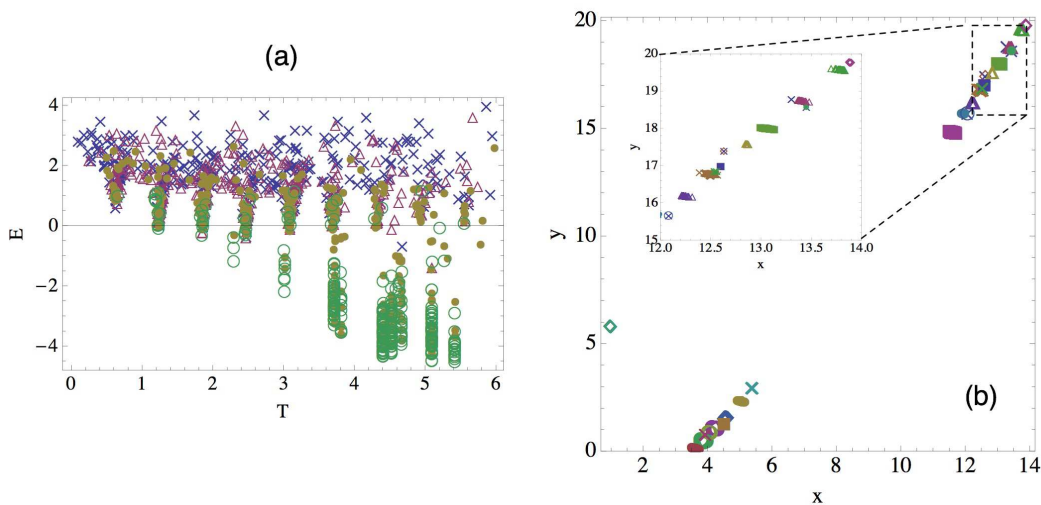}}
\caption{\label{fig:lorenz1} 
(a) Samples generated by the proposed method 
are plotted on the $(E,T)$ plane.  
The crosses, triangles, filled circles, and circles 
correspond to $\beta=2, 3, 4, 5$, respectively.
(b)  Samples with the energy lower than $-1.5$ are plotted on the Poincar\'{e} section $(x,y,r-1)$. The clusters, each of which corresponds to different 
minima, are identified by a cluster analysis method and they are plotted by a different symbols. 25 clusters are shown. The inset is a magnification.
$\epsilon=10^{-3}, N_e^{min}=1$, and $N_e^{max}=5$.}
\end{center}
\end{figure*}

For sampling of initial condition $\vec{x}_0$ of the Lorenz model, we take account of the symmetry $(x,y)\leftrightarrow (-x,-y)$ of the equations.
We sample initial conditions from the Poincar\'{e} section $\vec{x}_0=(x_0,y_0, r-1)$, where $x_0 \in [0,20]$ and $y_0 \in [0,20]$.
The period $T$ of orbits is assumed to be 
in the interval $[0,6]$. 
Using the energy function~(\ref{eq:lorenz_energy2}), the states $X=(\vec{x}_0,T)$ are sampled by the proposed method. 31 replicas with $\beta=2.0+0.1 i, \ i=0,1,\dots,30$ are used.

The samples obtained by the proposed method 
are plotted on the $(E,T)$ plane in Fig.~\ref{fig:lorenz1}(a) for different temperatures, which are obtained in parallel
in a replica exchange Monte Carlo calculation.
For lower temperatures, the energy 
takes smaller values for specific values 
of the period $T$, which correspond to UPOs.

\begin{figure*}[ht]
\begin{center}
\resizebox{\figuresize}{!}{\includegraphics{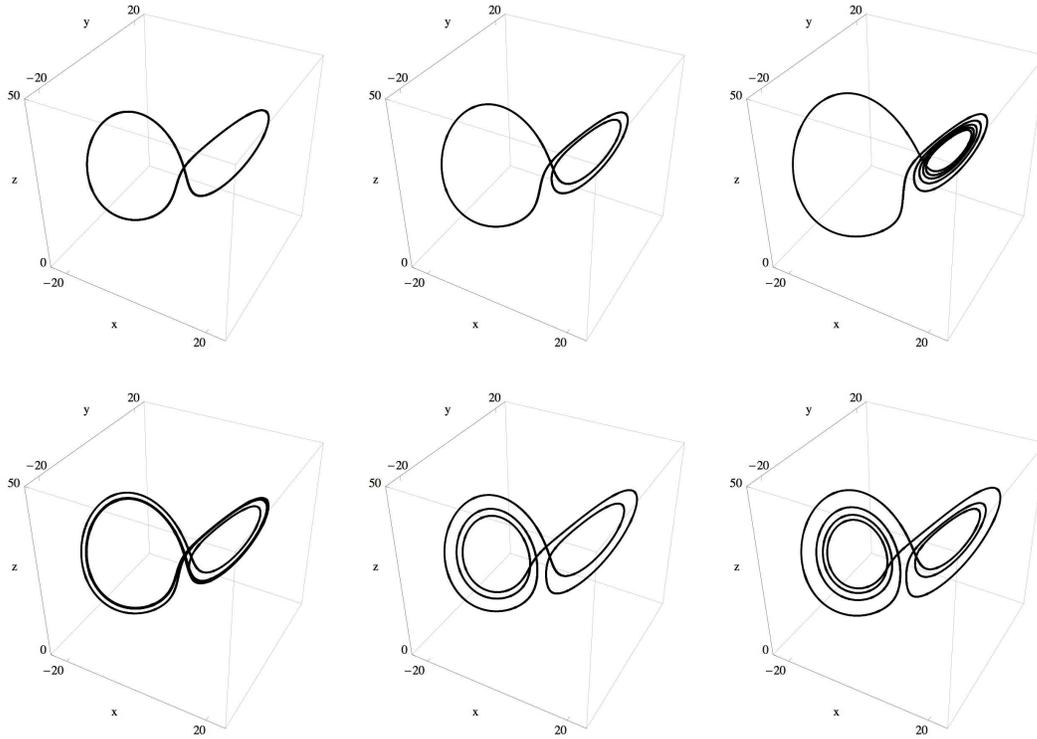}}
\caption{\label{fig:lorenz2} 
Typical unstable periodic orbits of the Lorenz 
model sampled by REM are shown. 
Each orbit is produced from 
time evolution starting from the initial condition that has the minimum energy in a cluster shown in Fig.~\ref{fig:lorenz1}(b). }
\end{center}
\end{figure*}

In Figure~\ref{fig:lorenz1}(b), the samples of initial conditions 
with $E(\vec{x}_0,T)\le -1.5$ are plotted on the Poincar\'{e} section.
The initial conditions form clusters and each cluster 
corresponds to initial condition in the vicinity of a different UPO.
In each cluster, we pick up the initial condition that has the minimum energy and calculate orbit $\varphi^T(\vec{x}_0)$ starting from the initial condition.
Typical orbits are shown in Fig.~\ref{fig:lorenz2}.
These orbits are closed in high precision, i.e.,  the 
difference between $\vec{x}_0$ and $\varphi^T(\vec{x}_0)$ 
is in order of $10^{-3}$, indicating that they are UPOs.

By using AUTO software\cite{AUTO}, we tested 
these orbits. An orbit detected by the proposed 
algorithm is used as an initial guess 
of a Newton's method implemented in AUTO.
We see that the convergence is remarkably quick, which indicates 
that the iteration begins ``sufficiently near'' 
UPOs. The difference between the output of the Newton's method 
and the initial guess is also very small.

%%%%%%%%%%%%%%%%% Pendulum %%%%%%%%%%%%%%%%%%%
\subsection{Stable Manifold of Unstable Fixed 
Points of a Double Pendulum}

Let us consider the following dynamics 
of the double pendulum with dissipation
\begin{eqnarray} \label{eq:doublep}
\fl \ddot{\theta _1}=\frac{\sin \left(2 \left(\theta _1-\theta _2\right)\right) \dot{\theta _1}^2+2 \sin
   \left(\theta _1-\theta _2\right) \dot{\theta _2}^2+3 \sin \left(\theta _1\right)+\sin
   \left(\theta _1-2 \theta _2\right)}{\cos \left(2 \left(\theta _1-\theta
   _2\right)\right)-3}  -k \dot{\theta _1} ,\nonumber \\
\fl \ddot{\theta _2}=-\frac{2 \sin \left(\theta _1-\theta _2\right) \left(2
   \dot{\theta _1}^2+\cos \left(\theta _1-\theta _2\right) \dot{\theta _2}^2+2 \cos \left(\theta
   _1\right)\right)}{\cos \left(2 \left(\theta _1-\theta _2\right)\right)-3}-k \dot{\theta _2},
\end{eqnarray}
where $k$ is a dumping coefficient. 

The double pendulum system~(\ref{eq:doublep})
has three unstable fixed points: 
$(\theta_1,\dot{\theta_1},\theta_2,\dot{\theta_2})=(\pi,0,\pi,0)$,
$(\pi,0,0,0)$ and $(0,0,\pi,0)$.  
Each unstable fixed point corresponds to an ``inverted'' 
state of pendulums.
Starting from any initial condition 
$\vec{x}_0=(\theta_1,\dot{\theta_1},\theta_2,\dot{\theta_2})=(0,\omega_1,0,\omega_2)$, however, almost all trajectories are converged to the 
stable fixed point 
$(\theta_1,\dot{\theta_1},\theta_2,\dot{\theta_2})=(0,0,0,0)$
by dissipation originate from the friction at the hinge.

In this system, we try to detect atypical 
trajectories that converge to unstable fixed points.
When an initial condition locates 
on the stable manifold of an unstable 
fixed point, the orbit starting from such 
an initial condition converges to the unstable 
fixed point after a long time evolution.
We search such an atypical initial state 
on the Poincar\'{e} section 
$(\theta_1,\dot{\theta_1},\theta_2,\dot{\theta_2})=
(0,\omega_1,0,\omega_2)$. 

\begin{figure*}[htb]
\begin{center}
\resizebox{\figuresize}{!}{\includegraphics{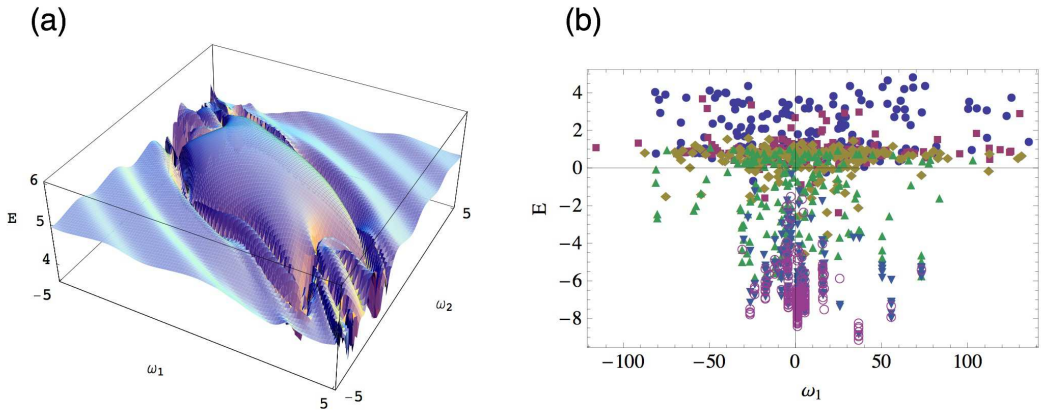}}
\caption{\label{fig:pen2-energy} 
(a) Rugged energy landscape $E(\vec{x}_0,T)$ with fixed $T=5.0$ as a function of $\omega_1$ and $\omega_2$.
(b) Samples are plotted on  $(E,\omega_1)$ for different temperatures; $\beta=0.1, 1.1, 2.1, 3.1, 4.1, 5.1$.
The sampling is performed by 51 replicas with $\beta=0.1 i$ for $i=1,2,\dots,51$.
$\epsilon=10^{-5}, N_e^{min}=1$, and $N_e^{max}=9$.
}
\end{center}
\end{figure*}

The state sampled by the proposed method 
is set as $X=(\omega_1,\omega_2,T)$.
There are many possibility of the energy function. Here we try to find trajectories converging one of the 
three fixed points and choose the energy as
\begin{eqnarray}
\fl E(\omega_1,\omega_2,T) & = & 
\log(E_k(\omega_1,\omega_2,T)+E_p(\omega_1,\omega_2,T)), \nonumber\\
\fl E_k(\omega_1,\omega_2,T) & = & 
|\dot{\theta}_1(T)|+ |\dot{\theta}_2^2(T)|, \nonumber \\
\fl E_p(\omega_1,\omega_2,T) & = & 
\min[\,\,\cos(\theta_1(T))\cos(\theta_2(T)), \,\,
\cos(\theta_1(T))+\cos(\theta_2(T))\,\,]+1,  \label{eq:pend} 
\end{eqnarray}
where $E_k$ represents ``kinetic energy'' of the 
pendulum at time $T$ stating from an 
initial condition defined by $(\omega_1,\omega_2)$.
The term $E_p$  represents an artificial ``potential energy'', 
which has minimum at each of the three unstable fixed points.
Note that if we want to find trajectories 
converging to a given unstable fixed point, 
we can use other artificial potential energies such as
$E_p(\omega_0,\omega_1,T)=|\theta_1(T)-\pi|+|\theta_2(T)-\pi|$.
The energy landscape defined by (\ref{eq:pend}) is shown 
in Fig.~\ref{fig:pen2-energy}(a), where the energy $E(\omega_1,\omega_2,T)$ 
vs. $\omega_1$ and $\omega_2$ is plotted for a fixed time, $T=5$.
Rugged structure is clearly seen and increase
of the sampling efficiency by REM is expected.

We discuss the result of a simulation with 
31 replicas with $\beta_i=2.0+0.2i$ for $i=0,1,\dots,30$.
In Fig.~\ref{fig:pen2-energy}(b), initial states sampled by the proposed 
method is plotted on the $(\omega_1,E)$ plane 
for replicas with inverse temperatures $\beta=0.1,1.1,2.1, 3.1, 4.1, 5.1$. 
The points are scattered at higher temperatures, 
but they are concentrated in separated regions 
when the temperature becomes lower.

\begin{figure*}[htb]
\begin{center}
\resizebox{\figuresize}{!}{\includegraphics{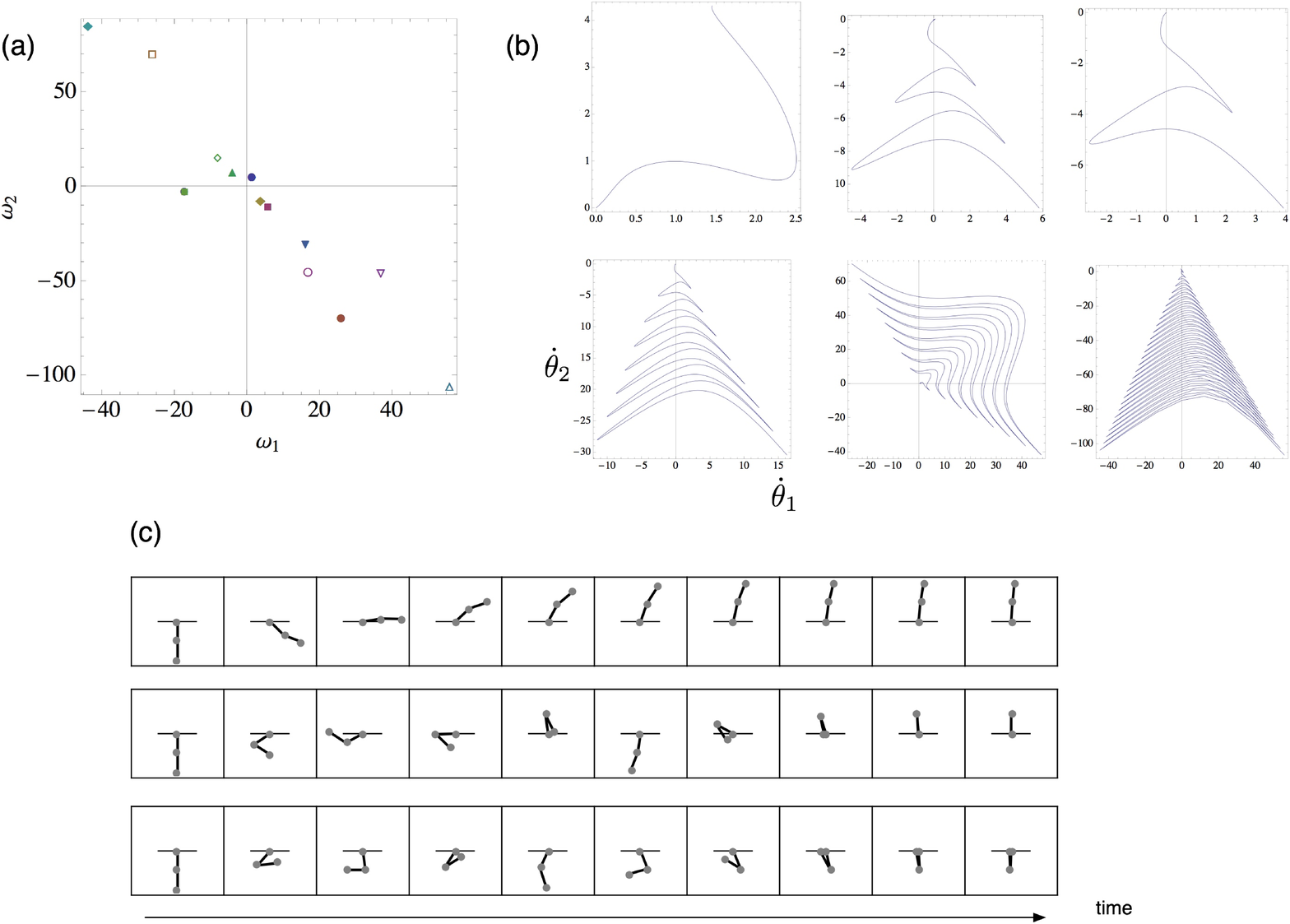}}
\caption{\label{fig:pen2-trajectory} 
(a) States with $E<-5$ is plotted on the $(\omega_1,\omega_2)$ plane. 
The states form clusters, which are regarded as sets 
of qualitatively different trajectories.
The different cluster is plotted with different symbols in 
the figure.
(b) Typical trajectories starting from the state that has minimum energy within a cluster are plotted on the $(\dot{\theta}_1,\dot{\theta}_2)$ plane.
Six trajectories are shown.
(c) Sequences of snapshots of pattern of the double pendulum that
correspond to trajectories.
Three trajectories, each of which converges to a different fixed point
are shown in the upper, middle, and lower panel.
The other parameters are the same as those in Fig.~\ref{fig:pen2-energy}
}
\end{center}
\end{figure*}

Because states with lower energies correspond to desired atypical 
trajectories, we select a set of the states whose energies $E$
are lower than a threshold $E_{cr}=-5$.
We show these states in Fig.~\ref{fig:pen2-trajectory}(a).
By using cluster analysis, we divide these initial states into clusters.
Since these clusters are well separated in the plane, 
we identify each cluster as a representation of a qualitatively 
different set of trajectories.

In each cluster, we identify an initial state that has minimum energy.
Figure~\ref{fig:pen2-trajectory}(b) demonstrates atypical 
trajectories starting from these initial conditions on $(\dot{\theta}_1,\dot{ \theta}_2)$ plane.  It is seen that each cluster corresponds 
to qualitatively different pattern of motion.
Examples of sequences of snapshots for pattern 
of the double pendulum are shown in Fig.~\ref{fig:pen2-trajectory}(c),
each of which corresponds to a trajectory obtained by the above
procedure.

%%%%%%%%%%%%%%%%%%%%%%%%%%%%%%%%%%%%%%%%%%
\section{Parameter Search and Combined Search}
\label{sec:extension}

In this section, search in a parameter space and 
combined search in parameter and initial condition spaces 
are discussed as an extension of our method. 
Examples are sampling of the boundary of the Mandelbrot set 
and exploration of the global bifurcation 
structure of the logistic map.

%%%%%%%%%%%%%%%%% Mandelbrot %%%%%%%%%%%%%%%%%%%

\subsection{The Boundary of the Mandelbrot Set}

For a map $f_c(z)=z^2+c$ parameterized by a parameter $c \in \mathbb{C}$, 
consider the sequence $(0,f_c(0),f_c(f_c(0)),\dots)$.
The Mandelbrot set~\cite{Mandelbrot04} is defined as the set of points 
$c \in \mathbb{C}$ such that the above sequence 
does not escape to the infinity. Hereafter, we denote 
$f_c(f_c(0))=f^2_c(0))$ and $f_c(f_c(f_c(0)))=f^3_c(0)$, and so on.

The Mandelbrot set is seen to have an elaborate boundary in the complex plane, 
which does not simplify at any magnification. 
This qualifies the boundary as a fractal.
In order to compute the boundary of the set, 
we use the parameter $c$ as a state of the proposed method
and employ the following energy function
$$
E(c)=-\log(n(c)),
$$
where the function $n(c)$ is defined as the smallest $n$
that $|f_c^{n}(0)| > 2$. If $n>N$ for a prescribed large
number $N=200$, we set $n(c)=1$ and $E(c) = 0$. 
In that case, the point $c$
is considered as inside the Mandelbrot set and not on the
boundary.

\begin{figure*}[ht]
\begin{center}
\resizebox{\figuresize}{!}{\includegraphics{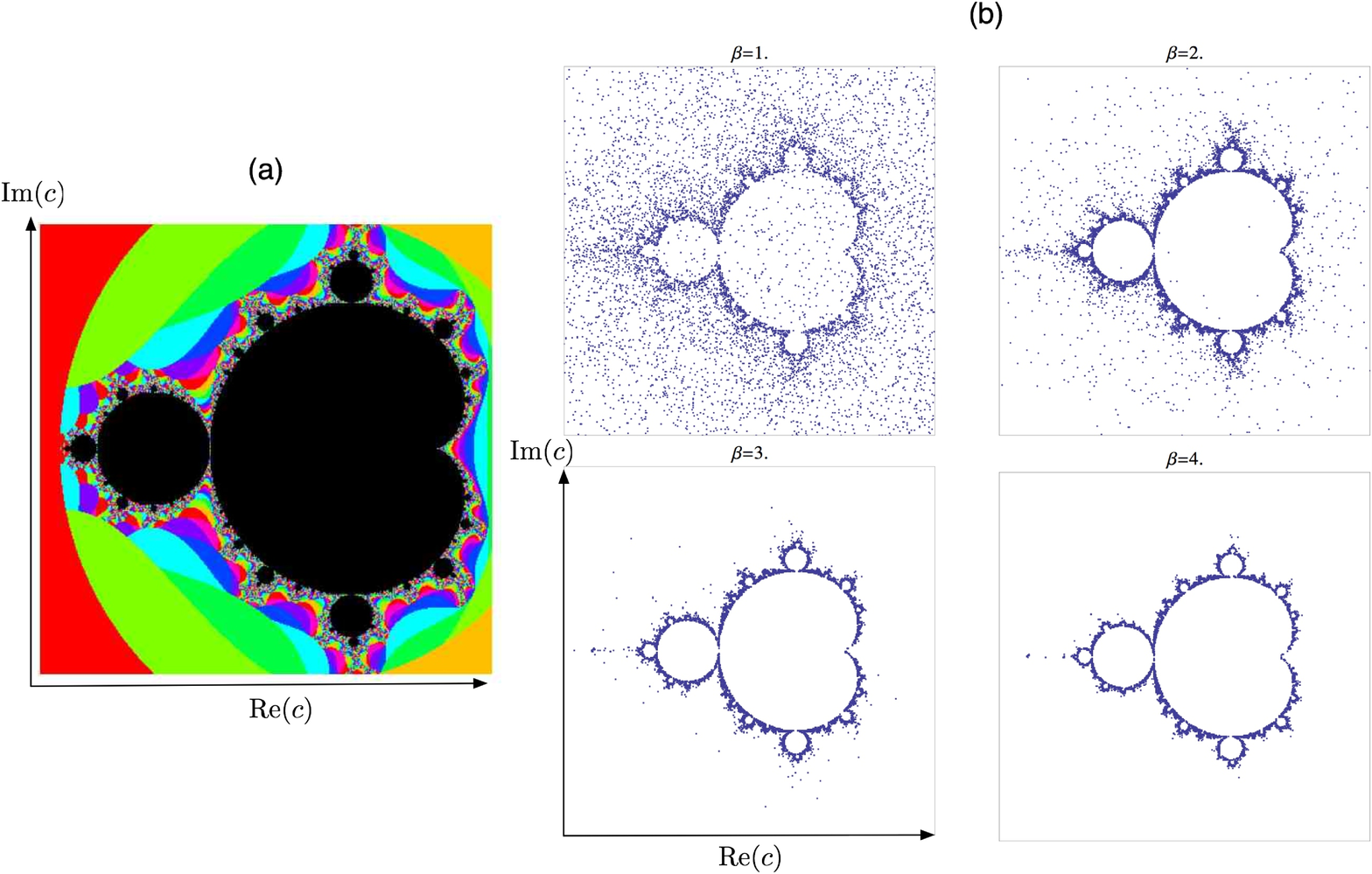}}
\caption{\label{fig:mandel} 
(a) The Mandelbrot set. The black region shows the Mandelbrot set and
the outside of the set is colored according to the escape time.
(b) The boundary of the Mandelbrot set.
Six replicas  with $\beta=1.0, 2.0, 3.0, 4.0 , 5.0,$ and $6.0$ are used for sampling.
$N_e^{min}=1$ and $N_e^{max}=9$.
}
\end{center}
\end{figure*}

Using this energy function, the boundary of the Mandelbrot set
 is calculated by  the proposed method  and the results are shown 
in Fig.~\ref{fig:mandel}. While the points are
almost randomly distributed in the complex plane 
for the replicas with a high temperature $1/\beta$,
the distribution of points corresponding to replicas with lower 
temperature is concentrated on the boundary of Mandelbrot set.

%%%%%%%%%%%%%%%%% Logistic %%%%%%%%%%%%%%%%%%%
\subsection{Periodic Orbits and Bifurcation 
Diagram of the Logistic Map}
\label{sec:logistic-periodic}
The logistic map \cite{May74,Ott02} is defined by
\begin{equation}
f_a(x)= ax(1-x),
\end{equation}
where $a \in [0,4]$ is a parameter.
Let us consider the energy function
\begin{equation}
E(a,x_0)=\log(|f_a^k(x_0)-x_0|+\epsilon), 
\label{eq:energy1}
\end{equation}
where $k$ is a given period and 
$\epsilon$ is a constant that determines the minimum energy.
Indeed, we can find initial conditions $x_0$ corresponding to 
period $k$ orbits for a given parameter $a$ 
using the energy function~(\ref{eq:energy1}) 
by the proposed method (the results are not shown here). 
On the other hand, 
the extension of sampling state space from $x_0$ to $(a,x_0)$
enables us the study of global bifurcation structure of 
the periodic solutions.
The extension is straightforward, i.e., we sample the vector $(a,x_0)$ 
from the Gibbs distribution determined by the same energy function 
(\ref{eq:energy1}) using REM. 

\begin{figure*}[ht]
\begin{center}
\resizebox{\figuresize}{!}{\includegraphics{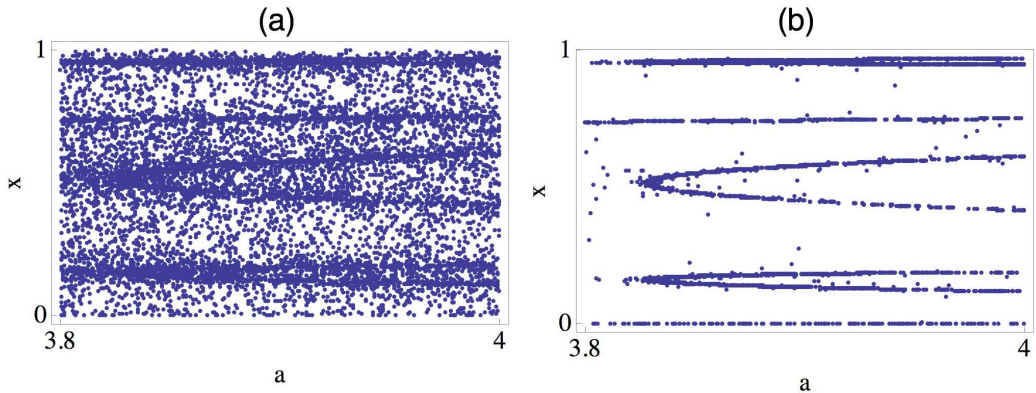}}
\caption{\label{fig:logistic} 
The bifurcation structure of the period 
three solutions of the logistic map is 
obtained by the proposed method.
Ten replicas are used with $\beta=0.5 i$ for $i=1,2,\dots,10$.
$\epsilon=10^{-4}, N_e^{min}=1$, and $N_e^{max}=11$.
(a) $\beta=0.5$.
(b) $\beta=1.0$.
}
\end{center}
\end{figure*}

In Fig.~\ref{fig:logistic}, the result for the period 
$k=3$ is shown, where
sampled points are plotted on the $(a,x_0)$ 
plane. 
We sample orbits from $a \in [3.8,4]$ and $x \in [0,1]$.
The points are scattered broadly in the vicinity of the true 
bifurcation branch of the period $k=3$ orbits at higher temperatures.
The dispersion of the points becomes smaller at lower temperatures and
periodic orbits with $k=3$ are detected.

%%%%%%%%%%%%%%%%%%%%%%%%%%%%%%%%%%%%%%%%%%%%%%%
\section{Summary and Discussion}

In this paper, we presented a general strategy
for exploring unstable
structures generated by nonlinear dynamical systems.
An artificial ``energy''
is defined as a function of the initial condition of the 
trajectory and the Gibbs distribution induced 
by the energy is sampled by the Metropolis method.
When an ``energy'' suitable for the purpose is chosen, 
sampling from the Gibbs distribution at low temperatures
realizes efficient sampling of rare initial
conditions that leads to interesting trajectories.
Replica exchange Monte Carlo (REM) is used
so as to avoid capturing in local minima 
of the energy landscape.

While implementation of the proposed 
method is simpler than the methods in which 
artificial energy is defined in the space of 
entire trajectories, the method can treat 
a variety of problem, including the search for 
unstable periodic orbits of the Lorenz model and stable
manifold of unstable fixed points of a double pendulum.

It is also shown that search in the parameter space
and combined search of initial conditions and parameters
become possible by adding the parameters to the state vector.
Two examples shown here are the Mandelbrot set of a complex
dynamical system and the bifurcation diagram of the logistic map.

An important future problem is to calculate quantitative
results by the proposed method. Examples are the relative 
densities of initial conditions that lead to different
Lyapunov numbers and densities of escape time from
a chaotic saddle. These calculations are possible
because REM is not only an optimization method, but also
realize correct sampling from the Gibbs distribution.
It is also interesting to introduce 
other types of extended ensemble Monte Carlo, such as
multicanonical algorithm, for this purpose. 
Research in this direction is
now in progress and the results will be published elsewhere.

The combined search
of initial conditions and parameters by the proposed method
is also promising and should be tested in the study 
of more complicated systems.
There is, however, an inherent interpretation
problem, i.e., the joint density in the direct product of 
parameter space and initial condition space seems not to have a
definite interpretation. The method is already useful to 
give a rough sketch of the bifurcation diagram, but it will be better
if we can give a physical meaning to the joint density.

%%%%%%%%%%%%%%%%%%%%%%%%%%%%%%%%%%%%%%%%%%%%%%%
\section*{Acknowledgments}
This study has been partially supported by the Ministry of Education, Science, Sports and Culture, Grant-in-Aid for Scientific Research (19540390).

%%%%%%%%%%%%%%%%%%%%%%%%%%%%%%%%%%%%%%%%%%%%%%%
\section*{References}
\bibliographystyle{unsrt}
\bibliography{jstat}

\begin{thebibliography}{10}

\bibitem{Ott02}
E.~Ott.
\newblock {\em Chaos in Dynamical Systems}.
\newblock Cambridge University Press, 2002.

\bibitem{CvitanovicWebBook}
P.~Cvitanovi\'{c}, R.~Artuso, R.~Mainieri, G.~Tanner, G.~Vattay, N.~Whelan, and
  A.~Wirzba.
\newblock {\em Chaos -- Classical and Quantum}.
\newblock \mbox{\tt http://chaosbook.org/}, unpublished (web book),.

\bibitem{Landau05}
D.~Landau and K.~Binder.
\newblock {\em A Guide to Monte Carlo Simulations in Statistical Physics}.
\newblock Cambridge University Press, 2005.

\bibitem{Newman99}
M.~E.~J. Newman and G.~T. Barkema.
\newblock {\em Monte Carlo Methods in Statistical Physics}.
\newblock Oxford University Press, 1999.

\bibitem{Liu01}
J.~Liu.
\newblock {\em Monte Carlo Strategies in Scientific Computing}.
\newblock Springer, 2001.

\bibitem{Young97}
A.~P. Young, editor.
\newblock {\em Spin Glasses and Random Fields}.
\newblock World Scientific, 1997.

\bibitem{Janke08}
W.~Janke, editor.
\newblock {\em Rugged Free Energy Landscapes: Common Computational Approaches
  to Spin Glasses, Structural Glasses and Biological Macromolecules}.
\newblock Lect. Notes Phys. Vol. 736. Springer, Berlin, 2008.

\bibitem{Mitsutake01}
A.~Mitsutake, Y.~Sugita, and Y.~Okamoto.
\newblock Generalized-ensemble algorithms for molecular simulations of
  biopolymers.
\newblock {\em Biopolymers (Peptide Science)}, 60:96--123, 2001.

\bibitem{Cho94}
D.~L.~Freeman A.~E.~Cho, J. D.~Doll.
\newblock The construction of double-ended classical trajectories.
\newblock {\em Chemical Physics Letters}, 229:218--224, 1994.

\bibitem{Bolhuis98}
Peter~G. Bolhuis, Christoph Dellago, and David Chandler.
\newblock Sampling ensembles of deterministic transition pathways.
\newblock {\em Faraday Discuss.}, 110:421--436, 1998.

\bibitem{Vlugt00}
T.J.H. Vlugt and B.~Smit.
\newblock On the efficient sampling of pathways in the transition path
  ensemble.
\newblock {\em Phys. Chem. Comm.}, 2:Art. No. 2, 2000.

\bibitem{Kawasaki05}
M.~Kawasaki and S.I. Sasa.
\newblock Statistics of unstable periodic orbits of a chaotic dynamical system
  with a large number of degrees of freedom.
\newblock {\em Phys. Rev. E}, 72:037202, 2005.

\bibitem{Sasa06}
S.I. Sasa and K.~Hayashi.
\newblock Computation of the {K}olmogorov-{S}inai entropy using statistitical
  mechanics: Application of an exchange {M}onte {C}arlo method.
\newblock {\em Europhys. Lett.}, 76:156--162, 2006.

\bibitem{Giardin06}
C.~Giardin\'{a}, J.~Kurchan, and L.~Peliti.
\newblock Direct evaluation of large-deviation functions.
\newblock {\em Phys. Rev. Lett.}, 96:120603, 2006.

\bibitem{Tailleur07}
Julien Tailleur and Jorge Kurchan.
\newblock Probing rare physical trajectories with {L}yapunov weighted dynamics.
\newblock {\em Nature Physics 3}, 3:203--207, 2007.

\bibitem{Bolhuis02}
P.~G. Bolhuis, D.~Chandler, C.~Dellago, and P.~Geissler.
\newblock Transition path sampling: Throwing ropes over mountain passes, in the
  dark.
\newblock {\em Ann. Rev. of Phys. Chem.}, 59:291--318, 2002.

\bibitem{vanErp07}
T.~S. van Erp.
\newblock Reaction rate calculation by parallel path swapping.
\newblock {\em Phys. Rev. Lett.}, 98:268301, 2007.

\bibitem{Doucet01}
A.~Doucet, N.~De Freitas, and N.~Gordon, editors.
\newblock {\em Sequential {M}onte {C}arlo Methods in Practice}.
\newblock Springer, 2001.

\bibitem{Iba01p}
Y.~Iba.
\newblock Population {M}onte {C}arlo algorithms.
\newblock {\em Transactions of the Japanese Society for Artificial
  Intelligence}, 16:279--286, 2001.

\bibitem{Kalos62}
M.~H. Kalos.
\newblock {M}onte {C}arlo calculations of the ground state of three- and
  four-body nuclei.
\newblock {\em Phys. Rev.}, 128:1791 -- 1795, 1962.

\bibitem{Sweet01}
D.~Sweet, H.~E. Nusse, and J.~A. Yorke.
\newblock Stagger-and-step method: Detecting and computing chaotic saddles in
  higher dimensions.
\newblock {\em Phys. Rev. Lett.}, 86:2261--2264, 2001.

\bibitem{Metropolis53}
N.~Metropolis, A.W. Rosenbluth, M.N. Rosenbluth, A.H. Teller, and E.~Teller.
\newblock Equations of state calculations by fast computing machines.
\newblock {\em J. Chem. Phys.}, 21:1087--1091, 1953.

\bibitem{Hukushima96}
K.~Hukushima and K.~Nemoto.
\newblock Exchange {M}onte {C}arlo method and application to spin glass
  simulations.
\newblock {\em J. Phys. Soc. Jpn.}, 65:1604--1611, 1996.

\bibitem{Iba01}
Y.~Iba.
\newblock Extended ensemble {M}onte {C}arlo.
\newblock {\em Int. J. Mod. Phys. C}, 12:623--656, 2001.

\bibitem{Biham89}
O.~Biham and W.~Wenzel.
\newblock Characterization of unstable periodic orbits in chaotic attractors
  and repellers.
\newblock {\em Phys. Rev. Lett.}, 63:819 -- 822, 1989.

\bibitem{Diakonos98}
F.~K. Diakonos, P.~Schmelcher, and O.~Biham.
\newblock Systematic computation of the least unstable periodic orbits in
  chaotic attractors.
\newblock {\em Phys. Rev. Lett.}, 81:4349 -- 4352, 1998.

\bibitem{Davidchack99}
R.~L. Davidchack and Y.-C. Lai.
\newblock Efficient algorithm for detecting unstable periodic orbits in chaotic
  systems.
\newblock {\em Phys. Rev. E}, 60:6172 -- 6175, 1999.

\bibitem{Lan04}
Y.~Lan and P.~Cvitanovi\'{c}.
\newblock Variational method for finding periodic orbits in a general flow.
\newblock {\em Phys. Rev. E}, 69:016217, 2004.

\bibitem{Lorenz63}
E.N.Lorenz.
\newblock Deterministic non-periodic flow.
\newblock {\em J. Atm. Sci.}, 20:130--140, 1963.

\bibitem{Sparrow82}
C.~Sparrow.
\newblock {\em The Lorenz Equations: Bifurcations, Chaos, and Strange
  Attractors}.
\newblock Springer-Verlag, New York, 1982.

\bibitem{AUTO}
E.~J. Doedel, R.~C. Paffenroth, A.~R. Champneys, T.~F. Fairgrieve, Yu.~A.
  Kuznetsov, B.~Sandstede, and X.~Wang.
\newblock Auto 2000: Continuation and bifurcation software for ordinary
  differential equations (with homcont).
\newblock {\em Technical Report, Caltech}, 2001.

\bibitem{Mandelbrot04}
B.~B. Mandelbrot.
\newblock {\em Fractals and Chaos: The {M}andelbrot Set and Beyond}.
\newblock Springer, 2004.

\bibitem{May74}
R.May.
\newblock Biological populations with nonoverlapping generations: Stable
  points,stable cycles,and chaos.
\newblock {\em Science}, 186:645, 1974.

\end{thebibliography}
% \label{}

%\begin{thebibliography}{00}

% \bibitem{label}
% Text of bibliographic item

% notes:
% \bibitem{label} \note

% subbibitems:
% \begin{subbibitems}{label}
% \bibitem{label1}
% \bibitem{label2}
% If there is a note, it should come last:
% \bibitem{label3} \note
% \end{subbibitems}

%\end{thebibliography}

\end{document}